\documentclass[10pt, conference, compsocconf]{IEEEtran}
\ifCLASSINFOpdf
  \usepackage[pdftex]{graphicx}
\else
\fi
\usepackage[cmex10]{amsmath}
\usepackage{color}
\usepackage{soul}

\begin{document}
%
\title{Multi-modal image retrieval with random walk on multi-layer graphs}



\author{\IEEEauthorblockN{Renata Khasanova\IEEEauthorrefmark{1},
Xiaowen Dong\IEEEauthorrefmark{2} and
Pascal Frossard\IEEEauthorrefmark{1}}
\IEEEauthorblockA{\IEEEauthorrefmark{1}Signal Processing Laboratory (LTS4),\\
EPFL,
Lausanne, Switzerland\\ Email: renata.khasanova@epfl.ch, pascal.frossard@epfl.ch}
\IEEEauthorblockA{\IEEEauthorrefmark{2}Media Lab, MIT, Cambridge, USA\\
Email: xdong@mit.edu}}


%


\definecolor{blue}{rgb}{0,0,0
}
\definecolor{red}{rgb}{0,0,0}

\maketitle

\begin{abstract}
The analysis of large collections of image data is still a challenging problem due to the difficulty of capturing the true concepts in visual data. The similarity between images could be computed using different and possibly multimodal features such as color or edge information or even text labels. This motivates the design of image analysis solutions that are able to effectively integrate the multi-view information provided by different feature sets. We therefore propose a new image retrieval solution that is able to sort images through a random walk on a multi-layer graph, where each layer corresponds to a different type of information about the image data. We study in depth the design of the image graph and propose in particular an effective method to select the edge weights for the multi-layer graph, such that the image ranking scores are optimised. We then provide extensive experiments in different real-world photo collections, which confirm the high performance of our new image retrieval algorithm that generally surpasses state-of-the-art solutions due to a more meaningful image similarity computation. 
\end{abstract}

\begin{IEEEkeywords}
Image retrieval, multi-modal data analysis, multi-layer graphs.

\end{IEEEkeywords}

%
\IEEEpeerreviewmaketitle

\section{Introduction}
Image collections, stored on the Internet, are rapidly growing every day. In the same time, retrieving relevant information in these huge data collections is an increasingly important challenge. Image retrieval systems can be constructed in different ways, where the user can provide the system with either a query image, text or characteristic for the search. We are interested in the case where the user provides a query image, and the task is to reorder the images in a multi-view dataset according to their relevance or similarity with the query image. We therefore develop a new query-based image retrieval algorithm, which has access to image labels for a part of the dataset.

The image similarity is usually estimated through the comparison of different features that characterize the images. However, extracting features that can effectively describe every image in the database is a challenging problem. This is often referred to as the semantic gap problem, as numerical features that we can extract from images cannot really describe the semantic information of the images. To tackle this, we propose to use multi-modal data gathered from different sources and to combine these multi-view features to compute the similarity between images. For example, textual and visual features can both describe an image and characterize diverse properties that are complementary to each other. However, these features can be fundamentally different: some features measured with real numbers (e.g., gradient histogram of an image \cite{bb:hog}) others are binary (e.g., indicator that shows that an image contains a particular object \cite{bb:bin_brief}). The effective combination of these different features for image retrieval is therefore a challenging problem.

\begin{figure}[!t]
\centering{
\includegraphics[height=1.8in]{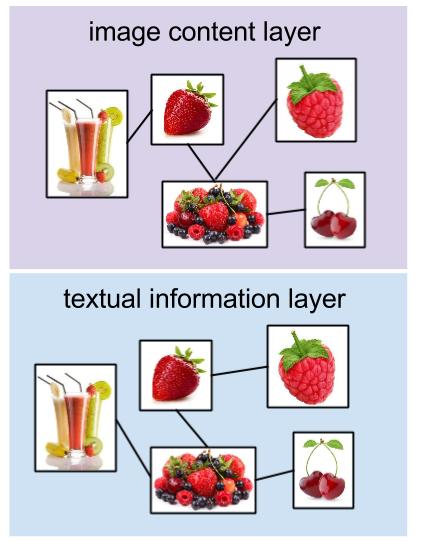}
\caption{Example of a multi-layer image graph. Each image is a node, and the edges in each layer depend on the similarity between images for a given type of feature.}
\label{fig:multi_layer}
}
\end{figure}


We propose in this paper a new algorithm based on multi-layer graphs to rank images in combining heterogeneous features. First, to work with big data collections, we offer to store relationship pairs between features of all images in a flexible graph structure (see Fig.\ref{fig:multi_layer}) where each type of modality forms a different layer. Using this structure, we preserve important information about each type of the feature. Second, we develop a random walk process on the graph to retrieve similarity information between images. This permits to cope with possibly incomplete data in some modalities. Our algorithm makes transitions between layers based on categories' distribution of current node's  neighborhood. It forms a flexible framework where labeled data can be used to give different node-specific weights to different layers. A detailed analysis shows that our algorithm converges to a stationary state that can be achieved using the same random walk process on the one-layer graph that is obtained by weighted node-specific summation of all graph's layers. Thus, our method can be seen as a generalization of a random walk algorithm to graphs constructed on multimodal features. Extensive experiments finally show that the proposed algorithm outperforms baseline and state-of-the-art methods in multimodal image retrieval.

In summary, our contribution is three-fold:
\begin{enumerate}
\item we develop a new image retrieval algorithm that works with possibly incomplete multi-modal features;
\item we propose a generalization of a random walk algorithm to multi-layer graphs;
\item our method is capable of adjusting the transition probabilities to the priority of different modalities.
\end{enumerate}

The paper is organized as follows. In Section \ref{sec:rel_work}, we describe the related work. The proposed algorithm for image retrieval task and the convergence analysis are given in Section \ref{sec:alg}, where we also describe an approach to find a transition probability matrix across layers. Experiments are discussed in Section \ref{sec:exp} and we conclude in Section \ref{sec:con}.

\section{Related Work}
\label{sec:rel_work}
An image retrieval algorithm obtains a ranking result using features that can distinguish the images from each other. These features can be divided into local and holistic ones. The former describe interest points in an image and the latter describe global characteristics of the image. In many cases the query image is described as a combination of features of different nature, thus, multi-modal fusion of the features is needed to respond to user queries \cite{bb:ImageRetrievalSurvey}.  The algorithms that use multi-modal data, can be divided into early and late fusion algorithms. In early fusion models, multi-modal data is combined at the feature level, while in late fusion they are rather combined at the output level.

There is a lot of the research work that has been dedicated to early fusion methods. In \cite{bb:bronstein}, the authors assume that multi-modal data lies on manifolds that are embedded in high-dimentional spaces. They construct multiple graphs using different features. Afterwards, they propose to find a common Laplacian matrix for the graphs. Other researchers use joint matrix factorization to build a unified optimization algorithm \cite{bb:pf}. In \cite{bb:join_diag}, the authors formulate a nonnegative matrix factorization constraint for clustering tasks which penalizes solutions without consensus between different features. Another method uses canonical correlation analysis \cite{bb:can_cor} and projects the multi-modal data into a low-dimensional space to eventually work with the projected data. In late fusion methods, such as \cite{bb:late1} and \cite{bb:latefusionkumar}, different ranking results are obtained for different features, and authors propose effective rules to aggregate them. It is however challenging to fuse ranks that are obtained using different features, because top results can have only a few common images or even empty intersections. 

\textcolor{blue}{In \cite{bb:earlyvslate}, authors compare early and late fusion methods for semantic video analysis using the SVM algorithm. Early fusion faces a computational issue because the resulting input vector, which is obtained using feature concatenation, is of high dimension. Also, early fusion is a challenging task because all features should have a similar representation. The disadvantages of late fusion include its expensive computational cost, because the results should be obtained for all feature vectors separately, which implies that the algorithm is run multiple times. For the sake of completeness, we finally note that there also exist middle fusion methods that keep important information about every type of feature. For example, the co-training \cite{bb:cotr} and co-regularization \cite{bb:core} methods work with multi-modal features and obtain a unified result that combines different information sources.} \textcolor{red}{These approaches exploit the diversity of different modalities in a limited scenario, because they are able to find solution which is only present in all modalities}.

At the same time, there is nowadays a growing interest in graph-based algorithms in supervised, semi-supervised and unsupervised image clustering, retrieval and classification tasks \cite{bb:graph}. Clustering could be efficiently performed on graphs using spectral clustering, for example \cite{bb:tut_lux}. Traditionally, image retrieval is based on a search of pairwise distances of all the images, which translates into finding the most similar neighbours on graphs. However, in this case, some important information about the distribution of the features of all images in the database can be lost. Context properties or data models can help to preserve this global information. The authors of \cite{bb:graph_re_manifold}, for example, propose to improve the result of unsupervised image retrieval using a manifold structure. A random walk model is used in \cite{bb:graph}, which looks for a combination of the initialization of the graph, the type of the transition matrix, and the definition of the diffusion process, which gives the best retrieval result. The method in \cite{bb:rerank} finally reorders images, using content features the images that are initially ranked based of textual information. The authors propose to learn a graph for every feature individually based on query images. After that, these graphs are combined into a single graph structure and the images are reordered accordingly. Graphs also play an important role in classification. The main assumption in classification tasks is that similar objects tend to belong to the same class. A lot of works based on regularization theory search for the smoothest graphs' signal \cite{bb:class_reg,bb:transfer_learning} for proper classification. For example, the authors in \cite{bb:class_reg} interpret the labels as a signal on graphs and their classifier finds the smoothest graph signal. However, all these methods are designed for a single data modality, we on the other hand propose an algorithm that effectively combines data from different sources.

In summary, there are a lot of methods that try to solve the image retrieval problem. However, the challenging semantic gap problem is still not fully solved \cite{bb:ImageRetrievalSurvey}. To tackle the problem, we propose to use a flexible, sparse multi-layer graph structure (Fig. \ref{fig:multi_layer}). The graphs capture information about the distribution of the images in the database, and the proper combination of multi-modal data addresses the semantic gap problem. In particular, we develop a new framework that permits to effectively handle data that can be incomplete in some modalities and introduce node-specific weights to effectively combine features on graphs.

\section{Random walk for image retrieval}
\label{sec:alg}

\subsection{Multi-layer graphs}
Images can be compared with each other using similarity measures, which can be conveniently represented by sparse graph structures. A graph $(V, E, W)$ is defined by a set of nodes $V$, edges $E$ and edge weights $W$. Each node is associated with one image and each edge represents the relationship between two images. The weight of the edge expresses the image similarity that is measured with particular features.

Data around us can typically be represented by multi-modal information, where different kinds of information complete each other. We therefore use a multi-layer graph to combine data from different sources into one single structure. For example, we can construct a multi-layer graph using textual and image content information. 

Assume that the system contains images of a raspberry and a berry smoothie (Fig. \ref{fig:multi_layer}). These images are not connected on textual and content layer because they do not have common tags and look different. However, these images are related to each other. A multi-layer graph structure helps to find this relationship. If the database contains an image with a basket of berries connected to a raspberry image at the content layer (they can have common local-level features) and to the smoothie image at the textual layer (they have in common the tag ``berry"). The multi-layer graph is then able to establish a relationship between a raspberry and a smoothie, which is reasonable from a semantic viewpoint.
 
We extract features of a varied nature --  content features, features based on tags, meta-data features and so on -- and use these features to construct different layers. Each layer $l$ of multi-layer graph $(V, E_1, \dots, E_L, W_1, \dots, W_L)$ is a single graph $(V, E_l, W_l)$. All layers have the same set of nodes $V$ but with different edges and edge weights $(E_l, W_l)$ in each layer. For example, in Fig. \ref{fig:multi_layer}, a two-layer graph for an image dataset is constructed. The images are the same for all layers, but the relationships between these nodes represent similarities in terms of different features, namely, textual and visual content of an image.

\subsection{Random walk on a multi-layer graph}
The image retrieval problem can be solved via a random walk process on the graph. A random walk consists in a succession of random steps driven by transition probabilities that depend on edge weights. The most visited nodes get high rank values in the retrieval result. If the graph is properly constructed, similar images are connected by strong edges, which increases their probability of being visited by the random walk. 

We create a random walk process on a multi-layer graph that is constructed using multimodal information as indicated above. We suppose that similar images are connected to each other in one or more layers. For example, sea and lake images are connected at the texture layer, and images of food and restaurants are connected at the text layer. Then, we start a random walk process on the query image. On every step, the algorithm can walk within a layer of the graph or can make a transition to another layer of the graph based on the relative importance of the layers (Fig. \ref{fig:multilayer_with_notations}). This transition between layers extends classical random walk process to multi-layer graphs, in order to benefit from the availability of multi-modal information. 

More formally, we start the algorithm with a vector of ranking values for all the images nodes $r^{(0)}=\pi$, where $\pi=\{\pi_{1},\pi_{2},\ldots,\pi_{M}\}$ is a fixed distribution with $\pi_{q}=1$ for the query node and $0$ for other nodes and where $M$ is a number of images. We then perform a random walk on the multi-layer graph. At each time step, we consider two sub-steps. On the first sub-step, we choose the layer $l$ for the node $i$ to perform the random walk with the probability $\alpha_{li}$. Afterwards, we choose the neighbor node $j$ to perform a random walk step according to the transition probabilities in layer $l$. Accordingly, we iteratively update the rank vector till convergence in the following way:

\begin{figure}[!t]
\centering{
\includegraphics[width=1.4in]{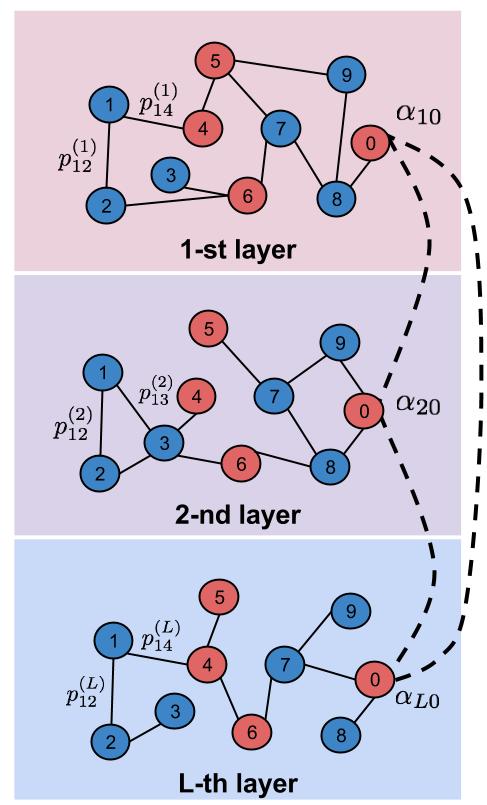}
\caption{Example of L-layer graph structure with labeled and unlabeled nodes in red and blue, respectively. On node $0$, we show the transition probability $\alpha_{l0}$ to choose layer $l$ to continue the random walk, and, on node $1$, we show the transition probability $p_{1j}(l)$ to make a random walk step toward the neighbor node $j$ on layer $l$ (best seen in color).}
\label{fig:multilayer_with_notations}
}
\end{figure}

\begin{equation}  
r^{(t)}=(1-\eta) \pi + \eta (P_1^T \Lambda_1 + P_2^T \Lambda_2 + \dots + P_L^T \Lambda_L) r^{(t-1)},
\label{form:random_walk}
\end{equation}
where $r^{(t)}$ is a ranking value on iteration $t$, $1 - \eta$ is the probability of jumping back to the query vertex, $P_l^T$ is a transition matrix for layer $l$ and $\Lambda_l$ is a node-specific matrix that represents the probability to choose the layer $l$ in the random walk. The transitional probability between node $i$ and node $j$ for layer $l$ is defined as 
\begin{equation}
p_{ij}(l) =\frac{w_{i,j}(l)}{\sum_{j\in N_i(l)} {w_{i,j}(l)}},
\end{equation}
where  $w_{i,j}(l)$ is the weight of the edge between nodes $i$ and $j$ for layer $l$, $N_i(l)$ is a set of the vertices that are the neighbors of vertex $i$ in layer $l$. The weights $w_{i,j}(l)$ simply represent the similarity between images $i$ and $j$ based on the features considered in layer $l$. These transitional probabilities $p_ij(l)$ that form the matrix $P_l^T$ can be calculated in a similar way for all the layers. 

Then, the layer transition probability matrices  $\Lambda_l$ for every layer $l$ are diagonal matrices $\Lambda_l=diag(\alpha_{l1}, \dots, \alpha_{lM})$, where $\alpha_{li}$ is a node-specific probability for jumping to layer $l$ at node $i$. Note that the sum of each rows across different layer's matrices is equal to $1$:

\begin{equation}
\sum_l^L \alpha_{li} = 1.
\end{equation}
We propose our method to compute these matrices in the next section.

\subsection{Layer transition probabilities}
\label{sec:nsp}

Choosing a layer for the random walk in a multi-layer graph is a process that has not been studied well. To the best of our knowledge, a method that uses node-specific probabilities to combine layers does not exist. 

We thus propose a method for computing the layer transition probability $\Lambda_l$. First, we observe that the layer transition probabilities could be different for different query images. For example, the textual layer is more important for query images from the ``animals" category, because it can connect an image of a ``cat" and an image of a ``cow", but the visual features layer is more important for categories ``sea" and ``sky", because it connects images with a common structure. Therefore, we suggest learning node-specific transition probabilities $\Lambda_l$ with respect to a query image. 

To calculate these probabilities, we assume that we know part of the labels in the image dataset. This means that we know the categories which some of the images belong to. Then, the idea is to favor a walk in the layer where most of the neighbors of the current node belong to the same category. For this purpose, we consider the labeled nodes in the neighborhood of a particular node $i$ within a radius $d_l$. The neighborhood of the node $i$ is a set of nodes, which are strongly connected to $i$. Nodes $i$ and $j$ are strongly connected if the weights $D_l(i,j)\geq d_l$, where $D_l(i,j)$ is defined as:
\begin{equation}
D_l(i, j)_{i,i_1, i_2\dots,j} = w_{i,i_1}(l)w_{i,i_2}(l) \dots w_{i,j}(l),
\end{equation}
with $\{i, i_1, i_2, \dots, j\}$ a path in a graph's layer. If there are several paths $k \in K$ between nodes $i$ and $j$ we choose the path that gives maximum weight value: 
 \begin{equation}
D_l(i, j) = \max_{k\in K}(D_l(i,j)_{k}).
\label{eq:distance}
\end{equation}
 
For example, let us assume that a graph has strong edges $e(i,j_1), e(j_1,j_2)$ and a weak edge $e(j_2, j_3)$. Then for the node $i$ the algorithm includes the nodes $j_1$ and $j_2$ to a neighborhood set, and stops to look for deeper neighbors for the node $i$. Thus, the neighborhood of each node contains only relevant neighbors for this node.

To calculate the layer transition probabilities, we then compute the number of categories in this neighborhood:
\begin{equation}
n(l,i)=\max_{c\in C}\frac{\#v_{li}(c)}{\#v_{li}},
\end{equation}
where $C$ is a set of categories, $\#v_{li}(c)$ is the total number of labeled neighbors for node $i$ in a layer $l$ that belongs to the category $c$, and $\#v_{li}$ is the number of labeled neighbors.

After that, we normalize the values $n(l,i)$ to sum to 1 for the node $i$ across all layers,  \textcolor{blue}{in order to calculate the probabilities $n^{(n)}(l,i)$.}  Then, since we want to find the images that are similar to the query node we prefer walking in a layer that is important for the query node with more probability than for the other layers. Therefore, the label distribution around the query image $q$ should also affect the transition probabilities. \textcolor{red}{Thus, we combine probabilities $n^{(n)}(l,i)$ and $n^{(n)}(l,q)$ and normalize the result so that the transition probabilities for a given vertex sum up to one.} We finally walk with the node-specific probability $\alpha$:
\begin{equation}
\alpha_{li} = \frac{z(l,i)z(l,q)}{\sum_l \big(z(l,q) z(l,i)\big)},
\label{eq:prob}
\end{equation}
where $z$ is a sigmoid function:
\begin{equation}  
z(l,i) = \frac{1}{1 + e^{-a(n^{(n)}(l,i) - n^*)}}.
\end{equation}

The sigmoid function gives higher priority to the probabilities that are larger than a threshold $n^*$, and lower priority to others, where $a$ is a coefficient that changes the sigmoid function's slope. The probability in Eq. \ref{eq:prob} is influenced by both the neighborhood of the node $i$ and the query $q$, through $z(l,i)$ and $z(l,q)$ respectively.

To sum up, we propose to calculate the layer transition probability for each node based on labeled nodes in its neighborhood. The neighborhood contains only nodes that have strong connections with each other and can vary from layer to layer. The algorithm gives more priority to a layer which contains many nodes from a similar category, because it is an indicator that the features in the layer properly represent this category.

\subsection{Convergence analysis}
\label{sec:prv}
We study now the convergence of the random walk proposed above. Since $(P_1^T \Lambda_1 + P_2^T \Lambda_2 + \dots + P_L^T \Lambda_L)$ is a large sparse matrix, we use the power method \cite{bb:poweralg} to calculate the equilibrium state without computing a matrix decomposition Eq. (\ref{form:random_walk}). The power method is an algorithm that iteratively applies to $r^{(t)}=Br^{(t-1)}$ and produces an approximation of an eigenvalue and an eigenvector for given matrix $B$ until convergence. We start from $r^{(0)}$, which has only one nonzero component that represents the query node. Finally, the retrieval system uses the ranking score computed after the convergence of an algorithm.

The method can be used when the transition matrix is column-stochastic and the distribution $\pi$ sums up to 1 \cite{bb:poweralg}, which is the case in the algorithm. According to the Perron-Frobenius theorem largest eigenvalue of the column-stochastic matrix is equal to $1$ and all the other eigenvalues have absolute value smaller than 1. Thus, if transition matrix has an eigenvalue, which is strictly greater in magnitude than the others, then the system converges to $r^*$ after several random walk steps \cite{Meyer00}, and we can write:




\begin{equation}
r^{*} = (1-\eta) \pi + \eta(P_1^T \Lambda_1 + \dots + P_L^T \Lambda_L) r^{*},
\end{equation}

or equivalently:
\begin{equation}
(I - \eta(P_1^T \Lambda_1+ \dots + P_L^T \Lambda_L)) r^{*}  = (1-\eta) \pi.
\end{equation}

Since we assume by the construction of the graph that there is at least one possible transition from any node to one of its neighbours at one of the layers, the matrix $(I - \eta(P_1^T \Lambda_1+ \dots + P_L^T \Lambda_L))$ is invertible. Hence we have that the solution after convergence can be written as:

\begin{equation}
r^{*} = (1-\eta) (I - \eta(P_1^T \Lambda_1+ \dots + P_L^T \Lambda_L)) ^{-1} \pi.
\end{equation}
This solution can also be found algebraically, however, computation of the inverse matrix requires $O(n^3)$ operations, thus in order to efficiently work with large matrices we use the power method \cite{bb:poweralg} to compute the solution of our problem.

Finally, we note the personalized PageRank algorithm \cite{bb:pr} that is similar to our algorithm but uses unified transition probability matrices converges to:
\begin{equation}
r^{*} = (1-\eta) (I - \eta P^T) ^{-1} \pi.
\end{equation}

An algorithm that uses equal probabilities to choose any layer, would then converge to:
\begin{equation}
r^{*} = (1-\eta) (I - \eta \frac{1}{L} \sum_{l=1}^L P_l^T) ^{-1} \pi.
\label{eq:equal}
\end{equation}

We can clearly see some analogy between the convergence points of the different algorithms.  \textcolor{blue}{Our algorithm with a specific way to combine layers converges to the same state as a single layer graph framework. However, using simple single-layer graph-based methods for image retrieval is not feasible, as it will require building different graphs for different query nodes, which is computationally expensive.} Therefore, our algorithm can be seen as an extension of the standard random walk algorithm, which however properly takes advantage of different modalities.

\section{Experiments}
\label{sec:exp}
We now propose extensive experiments to evaluate the performance of the proposed algorithm. First of all, we present the evaluation metrics, the details of the graph construction and the dataset under consideration. Then we study in detail the behavior of our algorithm and the influence of the design parameters. Finally, we provide comparative results with state-of-the-art algorithms.

\subsection{Experimental settings}
\subsubsection{Evaluation metrics}
The objective of our retrieval algorithm is to obtain a ranking of images, where all images in the first positions should have a similar category as the query node. Assume that we know the ground truth image categories for all our dataset. We can thus measure the quality of our result using the mean Average Precision function (mAP) that estimates the quality of ranking for different queries. The function calculates the average precision (APr) for all queries from a dataset. More formally, let $M$ denote the number of images that are relevant to the query image in a database of $N$ images. Let $I(k)$ be an indicator function, which is equal to $1$ if the item at position $k$ in the image ranking is a relevant image, and zero otherwise. Let further $Pr(k)$ be the precision of the top k-rank values. We can then define APr and mAP as:

\begin{equation}
APr(q)=\frac{\sum_{k=1}^{N}Pr(k)I(k)}{M},
\end{equation}

\begin{equation}
mAP=\frac{\sum_{q\in N}APr(q)}{|N|}.
\end{equation}
The mAP metric is calculated for the complete rank result. However, in many cases, only the top rank results are interesting for the users. Therefore, the average NDCG metric \cite{bb:NDCG} can be used instead to compare the proposed method with other algorithms. The NDCG measure for a particular query node $q$ is calculated in the following way:

\begin{equation}
NDCG@P(q)=Z_p\sum_{i=1}^P \frac{2^{l(i)-1}}{\log(i+1)},
\end{equation}
where $P$ is the depth that describes how many nodes from the top result are considered, $l(i)$ is the indicator that shows that the node in position $i$ is relevant, and $Z_p$ is a normalization constant that sets the ideal score of NDCG to $1$ when all images on top positions belong to the same category as the query image. 

\textcolor{blue}{For the dataset where only a few examples are available the  N-S score is used. It represents the average number of correct images from top $M$ retrieved images, where $M$ is the size of the ground truth data set}.

\subsubsection{Graph construction}
\label{sec:gc}
For all our experiments we construct graphs where the edge weights are computed using Gaussian kernels to emphasize larger similarity values. In each layer, we define the edge weight between the corresponding feature vector $x_i$ of image $i$ and the respective feature vector $x_j$ of image $j$ as

\begin{equation} 
\label{eq:gaussian_kernel}
w_{i,j} = \exp \bigg(\frac{-||x_i - x_j||^2}{\sigma^2}\bigg),
\end{equation}
where $\sigma$ is used to adjust the degree of similarity between nodes. In order to construct sparse layers, we connect a node only with its $k$ nearest neighbors (in terms of Euclidean distances). Our graph is therefore undirected.

\subsubsection{Datasets}
To evaluate the effectiveness of our method we consider several datasets, including: MIRFlickr-25000~\cite{bb:mir},  Holidays \cite{bb:holidays} and Ukbench \cite{bb:uk}. Each of them has corresponding ground-truth annotation.

The size of MIRFlickr is $25000$ images. The annotation for this dataset comprise $10$ categories and $19$ subcategories. The goal of image retrieval is to obtain a list where the top images belong to the same category as the query image. The work in \cite{bb:mir} provides image's tags, edge histogram descriptor (EHD \cite{bb:mir_features}) and homogeneous texture descriptor (HTD \cite{bb:mir_features}) features for this dataset. Then, the Holidays dataset contains $1491$ images where $500$ of them are query images. For every query image there is a groundtruth list of corresponding  relevant images. Finally, Ukbench dataset is released with $10200$ images that are grouped into $2550$ clusters. For every image three corresponding images are known and provide the groundtruth information.

For the MIRFlickr dataset we use the Color Descriptor Koen library \cite{bb:Koen} to calculate feature points of images in the dataset and their descriptors. 
In particular we use the Harris Laplace detector to find feature points. For each of these points we extract the following color descriptors: colormoments, huesift, nrghistogram, opponenthistogram, rgbhistogram, SIFT. For a more detailed explanation on how these descriptors are computed, please refer to [26]. Each of these descriptors represents a different data modality and is computed from image feature points and the Bag-Of-Words model. Within each modality this process results into every image having a feature vector.

For the Holidays and Ukbench datasets we use  HSV Histogram, Convolutional Neural Network, GIST and Random Projection features from work \cite{bb:cvpr_query}. Each of these features is used to construct a layer in our multi-view graph.


\subsection{Analysis of the algorithm}
We first analyse the influence of the design parameters on the performance of our algorithm and run experiments for MIRFlickr dataset with $5000$ images. To construct the graph, we connect every node with five of its neighbours in each layer and we compute the edge weight according to Eq.~(\ref{eq:gaussian_kernel}). We run a 5-Fold cross validation on MIR Flickr collectionto choose the parameters of our algorithm.

\subsubsection{Radius parameter}
The radius parameter $d_l$ is used to calculate the layer transition probability, as defined in Section \ref{sec:nsp}. In the every layer $l$ we relate the neighborhood radius with the mean weight degree values:

\begin{equation}
\bar{w}(l) = \frac{\sum_{ij, (i, j) \in E_l} w_{i,j}(l)}{|E_l|},
\label{eq:mean_w}
\end{equation}
where $|E_l|$ is the number of edges in the graph at layer $l$. We now run experiments with different factor $\beta$ that linearly relates the radius parameter $d_l$ using the mean weight degree $\bar{w}(l)$ for each layer $l$:
 
 \begin{equation}
d_l = \beta \bar{w_l}.
\end{equation}

We define the layer transition probabilities with the different values of $d_l$ according to the Section~{\ref{sec:nsp}}, we choose $500$ nodes as query images over $5000$ images. We run 5-fold cross validation of our approach where for every fold, the algorithm has access to $1000$ labeled nodes. The resulting NDCG scores for each value of the radius $d_l$ are given in Table~{\ref{tab:radius_fac}}. Note that we evaluate the NDCG scores only for the queries with the highest and the lowest AP values. The queries with the highest score are well represented by features and ``easier" to rank. For these queries the choice of a radius does not have much of an influence on the solution. However, for the queries with low precision small radius gives better results, as it is important to choose the layer for the random walk based on probabilities computed in a small neighborhood. Table \ref{tab:radius_fac} depicts this analysis. As a result, in our further experiments we set $\beta = 0.5$.



\begin{table}
\centering
\begin{tabular}{|c|c|c|c|c|}
  \hline
  $\beta$&5@NDCG&10@NDCG&20@NDCG&40@NDCG\\
  \hline
  $0.5$ & \bf{0.3885} & \bf{0.2888} & \bf{0.2139} & \bf{0.1478} \\
  \hline
  $1.5$ & $0.3392$ & $0.2745$ & $0.2035$ & $0.1370$ \\
  \hline
  $2.5$ & $0.3392$ & $0.2863$ & $0.2036$ & $0.1379$ \\
  \hline
  $3.5$ & $0.3392$ &$0.2752$ & $0.1988$ & $0.1362$ \\
  \hline
\end{tabular}
\caption{NDCG performance for ranking images with different values of the radius parameter. The NDCG metric queries is calculated only on the 20 lowest AP scores. }
\label{tab:radius_fac}
\end{table}

\subsubsection{Return probability (1-$\eta$)}
Recall that the random walk algorithm can jump back to the query node with the  probability $1-\eta$ (see Eq. \ref{form:random_walk}). By varying $\eta$, we influence the number of nodes that the random walk algorithm visits. This parameter ideally depends on different configurations of the graph. The idea is that the algorithm should visit only the nodes that belong to the same category as the query node. We run some experiments to test the influence of the parameter $\eta$ on the performance of the algorithm for real data from the MIRFlickr collection.

Our experiment shows that the optimal parameter $\eta$ is actually different for different query images. Intuitively, for queries with high AP score it is easy to find relevant images, because the database contains many of them and the graph represents well the relationship between these images. In this case the random walk should visit all of them, therefore $\eta$ should be high. It might not be the case though for the low AP score queries, as only the closest neighborhood of the query node contains relevant images. Our experiments confirm this hypothesis. We found that setting $\eta$ to $0.8$ or $0.9$ gives higher NDCG score for queries with high AP score, while NDCG is not very sensitive to the choice of $\eta$ for the queries with low AP score.  Thus we set $\eta=0.9$ in our further experiments.

\subsubsection{Amount of labeled data}
Our algorithm is semi-supervised and we assume that part of the labels is known for the algorithm to compute the layer transition probabilities. We study now the influence of the amount of labeled data on the performance of our algorithm. We run the experiments on the MIRFlickr collection with $\eta=0.9$ and observe that for $12\%$ of labeled data our algorithm achieves a score of $5@NDCG=0.92$ for queries with high AP metric value and $5@NDCG=0.36$ for low AP metric. For $33\%$ of labeled data, our algorithm achieves values of $0.93$ and $0.36$ respectively. Thus, the algorithm gives slightly better results when more labels are available for the queries with high precision, and the amount of labeled data does not have much of influence on the solution for queries with low precision. Thus, this experiment shows that $12\%$ of the labeled data is enough to achieve good result for the dataset.

\subsection{Comparison with state-of-the-art methods}
\subsubsection{Illustrative example}
\begin{figure}[!t]
\begin{minipage}[h]{1\linewidth} 
\center {
\includegraphics[width=1in]{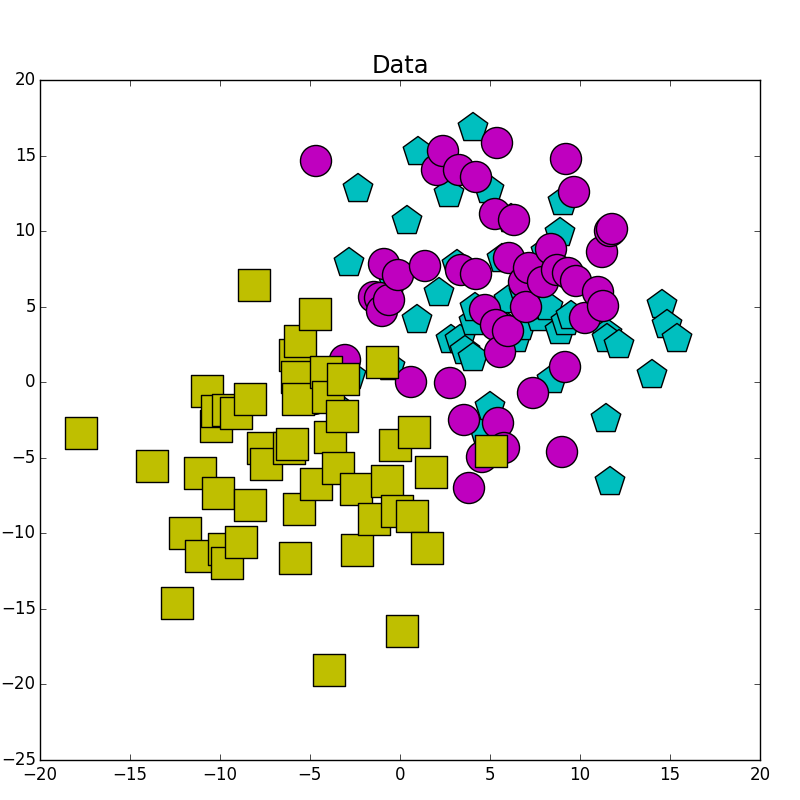}
\includegraphics[width=1in]{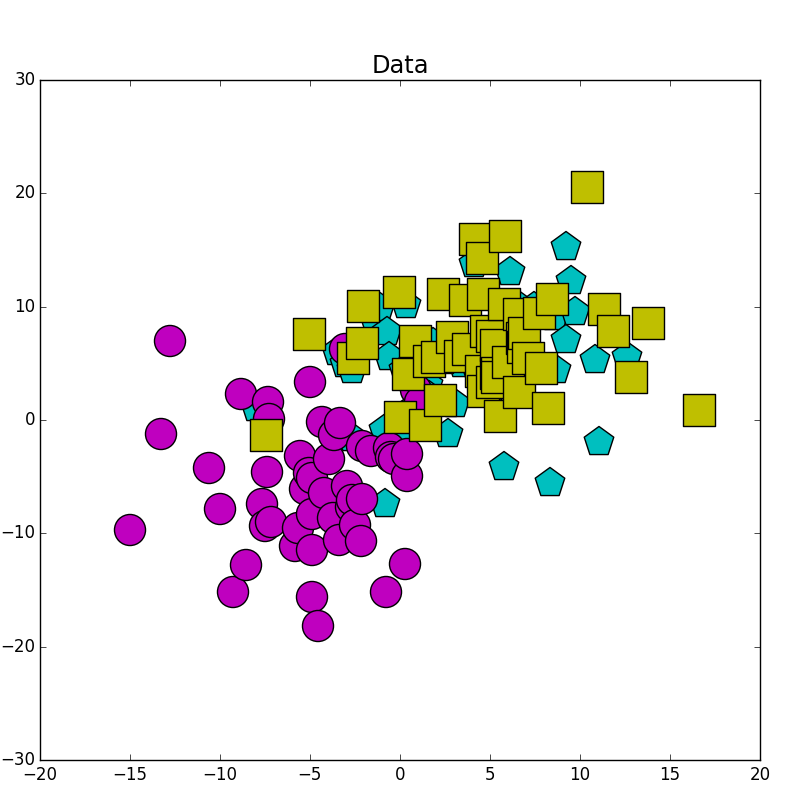}
\includegraphics[width=1in]{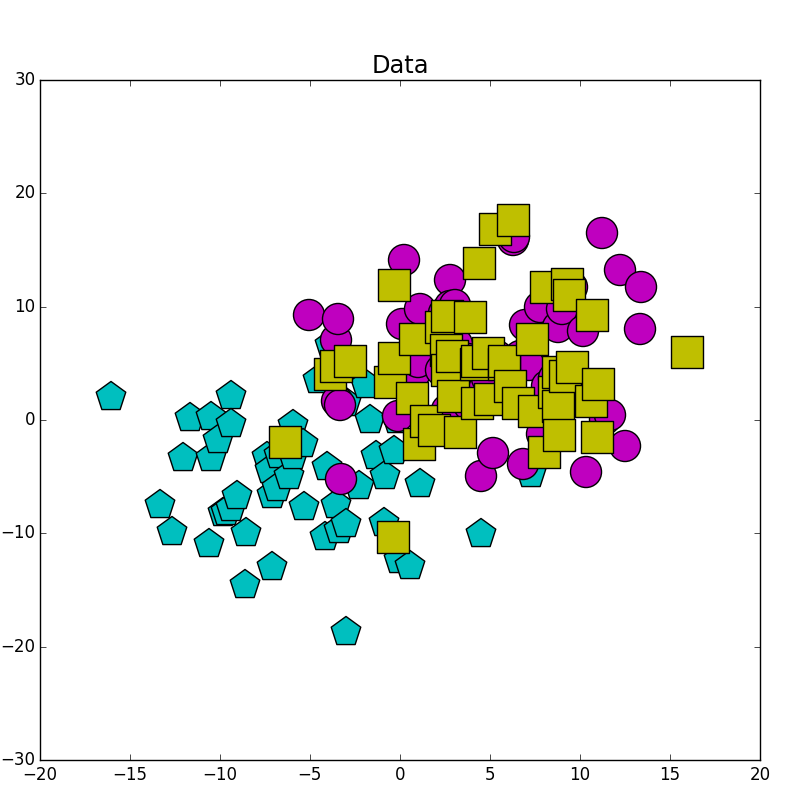}
} 
\end{minipage} 
\caption{Synthetic data that is represented by three layers. The color and shape show a category. The spatial coordinates represent the two dimensional features for the three layers respectively in each row of the figure (best seen in color).}
\label{fig:3layer}
\end{figure}

We first illustrate the behavior of our idea on a synthetic example (Fig.~\ref{fig:3layer}). Our synthetic data is represented by three layers, where each layer corresponds to different two-dimensional features. From Fig.~\ref{fig:3layer} we can see that every category is well separated only in one of the layers. \textcolor{blue}{We now compare our algorithm with a baseline one and with random walk algorithms \cite{bb:state_graph}. The baseline algorithm runs on multi-layer graph with $L$ layers and uses equal transition probabilities $\alpha_{l}=\frac{1}{L}$ for each node. The method in \cite{bb:state_graph} combines different layers of a multi-layer graph into one single graph. We show that our way of choosing layers according to layer-specific probability gives an advantage to our algorithm over the baseline and state-of-the-art methods. }

\begin{table}
\centering
\begin{tabular}{|l|c|c|c|}
  \hline
  Class label & Magenta  & Red  & Yellow\\
  \hline
  Baseline & 0.7271  & 0.6843 & 0.6930 \\
   \hline
  \cite{bb:state_graph} & 0.7355 & 0.7205 & 0.7249 \\  
   \hline
  Our algorithm & \bf{0.8502} & \bf{0.8617} & \bf{0.7287} \\  
  \hline
\end{tabular}
\caption{NDCG@10 metrics for synthetic data example of Fig. \ref{fig:3layer} with different probabilities selection methods}
\label{tab:3layer}
\end{table}

In particular, the average precision measure NDCG@10 of the ranking results of these random walk algorithms is given in Table \ref{tab:3layer}. Our method outperforms the baseline and the algorithm of \cite{bb:state_graph} for all data categories. For the first category our method achieves the best result because the features are well separated on the second layer while on the first and third layers they are mixed with objects from the other categories. Therefore, our method with node-specific transition probabilities across the layers chooses second layer with high probability for the random walk. This example shows that our approach outperforms the baseline probability selection method. It confirms the capabilities of our algorithm and motivates us to evaluate the method on real datasets.

\subsubsection{
Comparison with image reranking methods}


We now compare our retrieval method with the reranking algorithm \cite{bb:state_2} as both application share similarities, and our method is actually using a ranking score for retrieval. For the reranking algorithm \cite{bb:state_2} it is necessary to have an initial ranking result, which is obtained using textual information. Therefore, the algorithm performance can be evaluated only on a dataset which contains textual information. Afterwards, this algorithm reranks images from this initial set based on content features. \textcolor{blue}{The method in \cite{bb:state_2}, as our algorithm, uses labeled data to train the parameters.}


We conduct the experiments with only 5k images from 25k images of the MIRFlickr dataset \cite{bb:mir} due to the computational time of the methods under consideration. The dataset provides the following features: EHD, HTD and tags. \textcolor{blue}{We construct kNN graphs for EHD and HTD features with $k=15$ and $k=150$ respectively.} We use tag information to obtain the first approximation result for  \cite{bb:state_2} and to construct a graph layer for our method. To provide a fair comparison we consider only the image queries that contain tags \textcolor{blue}{that are shared with other images from the dataset} and include at least 100 images of the same category in their ground truth set. \textcolor{blue}{We chose 74 queries among the 5k images under consideration.}

\begin{figure}[!t]
\begin{minipage}[h]{1\linewidth} 
\center {
\includegraphics[width=2.4in]{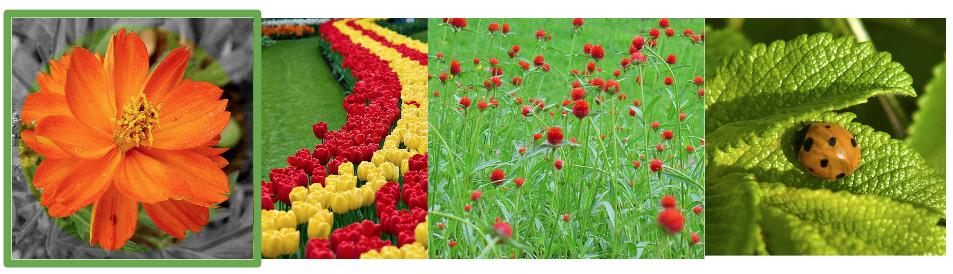}
} 
\center{a)}
\end{minipage} 
\vfill
\begin{minipage}[h]{1\linewidth} 
\center {
\includegraphics[width=2.4in]{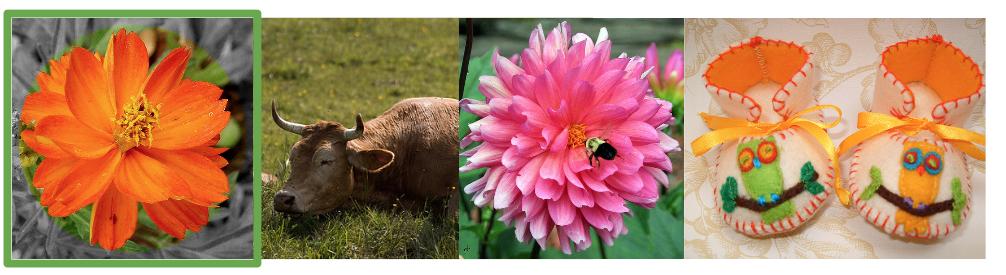}
} 
\center{b)}
\end{minipage} 
\caption{Top results obtained by a) our algorithm, b) \cite{bb:state_2}. The first image in each row is the query image. }
\label{fig:visual}
\end{figure}

\begin{figure}[!t]
\center {
\includegraphics[width=1.8in]{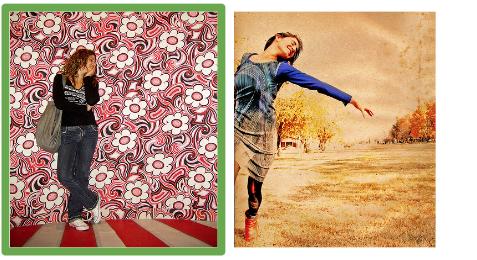}
} 
\caption{The left image is a query image and the right image is an image from the top $20$ retrieval results that is obtained by our algorithm. The right image does not contain any textual information. }
\label{fig:visual_2}
\end{figure}

\textcolor{blue}{Fig.~\ref{fig:visual} first illustrates the top results obtained by the proposed algorithm and the ranking method of \cite{bb:state_2}. The first image in each row of Fig.~\ref{fig:visual} is a query image. We can notice that the textual information permits to find images that are close to the query image in terms of semantic meaning but that look quite different usually.} Our algorithm gives a solution that combines the information from visual and semantic features. Furthermore, our method able to find images even without textual information, which can not be done using \cite{bb:state_2}. Fig.~\ref{fig:visual_2} illustrates this aspect: the relevant image shown on Fig.~\ref{fig:visual_2} is found among the top $20$ retrieval results with our algorithm, but not with \cite{bb:state_2}, since this image does not contain any tags. 


\begin{figure}[!t]
\center {
\includegraphics[width=3.6in]{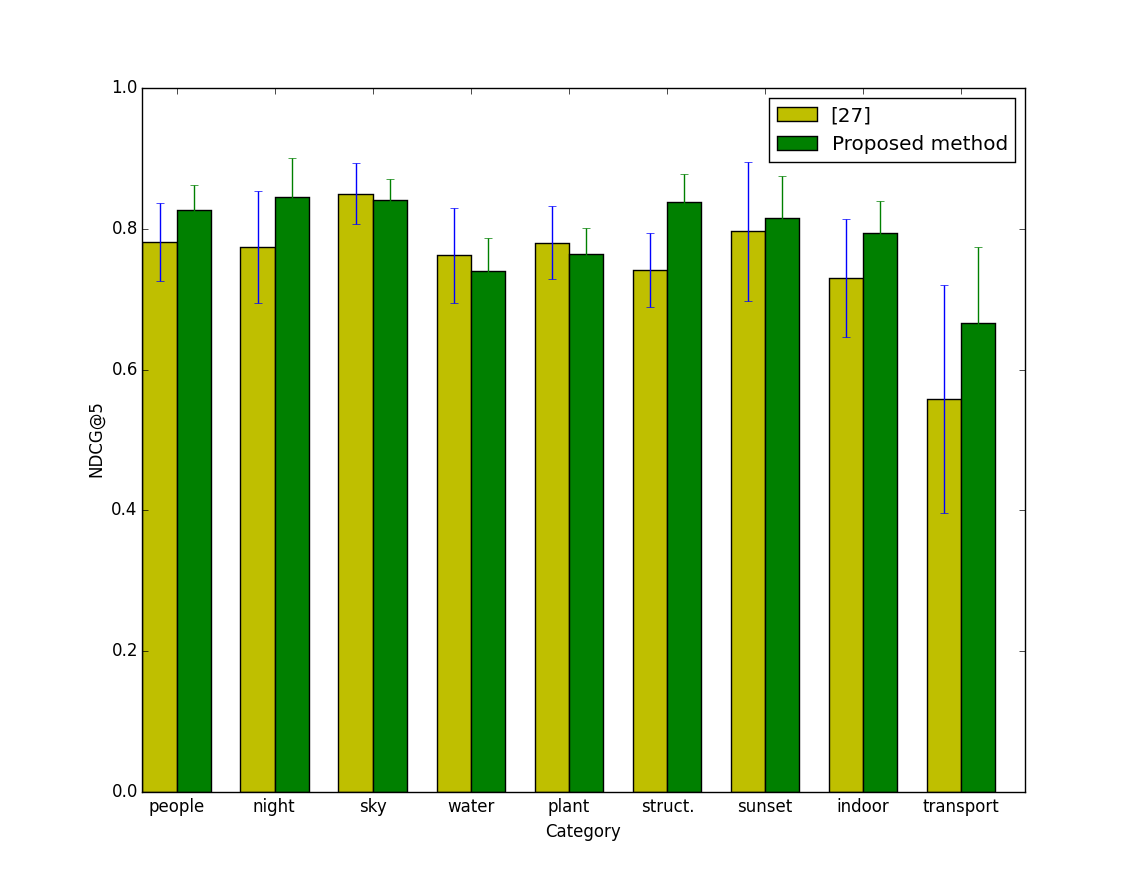}
} 
\caption{\textcolor{red}{Comparison of the NDCG@5 for our algorithm and \cite{bb:state_2} evaluated for different categories in the MIRFlickr collection.}}
\label{fig:NDCG5}
\end{figure}

\textcolor{blue}{We then compute NDCG metrics to compare the performance of the multi-modal reranking algorithm and our method. Since, in practice, the first images have usually a more important meaning than the others, we set the depth value to $N=5$. Fig. \ref{fig:NDCG5} shows the result of the corresponding experiment  for different categories. Our method outperforms \cite{bb:state_2} almost for each category. These algorithms achieve NDCG@5 values of $0.75$ and $0.70$ respectively. Also, the variance of the proposed algorithm is less than the one of the state-of-the-art method. Thus, our method is able to extract the most relevant images among the top results  and to achieve stable results}.


Finally, we run an experiment, which predicts the category of the images based on the ranking result. We organize it in the following way: we count the categories of the top $N$ results; based on this information, we assign the label of most frequent category to the query image. To evaluate the performance of different algorithms we check that the query image actually belongs to that category. The final prediction value is the probability of predicting the correct category for the query images. The best solution for this prediction is the following: for our random walk algorithm we are able to predict one of the query image category in $90\%$ cases using the top $10$ results, and for the reranking algorithm \cite{bb:state_2} in $80\%$ of the cases using $30$ results. This simple experiment further highlights the benefit of our algorithm in properly exploiting the multi-view features for image ranking.

\subsubsection{Comparison with image retrieval methods}
In this section, we finally compare our and state-of-the-art \cite{bb:cvpr_query} algorithms, which tackles multi-modal image retrieval problems using a late fusion strategy, and \cite{bb:state_graph}, which uses an early fusion strategy on the Holiday and Ukbench datasets. In \cite{bb:state_graph}, the authors aggregate the layers of a multi-layer graph into one layer. \textcolor{blue}{Notice that our results can be slightly different from the ones actually reported in \cite{bb:cvpr_query}, \cite{bb:state_graph}, as we do not use the exact same set of features as in these papers due to high complexity of extraction of some of them. In our experiments, we however use the same set of features for all algorithms under comparison.}

We also compare our algorithm with the following baseline algorithms:
\begin{itemize}
\item \textbf{Baseline 1.} Random walk with the equal transition probabilities  $\alpha_l = \frac{1}{L}$ for all graph's nodes and graph layers, where $L$ is a number of layers.
\item \textbf{Baseline 2.} \textcolor{blue}{To justify the node specific probabilities we compare our method with} a random walk with the equal transition probabilities $\alpha_{lq}$ for all graph's nodes but the choice of this probability is individual for every query and layer. These probabilities are calculated in the same way as proposed in our algorithm accordingly to Eq.~(\ref{eq:prob}) but only for the query node.
\item \textbf{Baseline 3.} We combine all features into one single vector and sort images according to their distances to the query image. 
\end{itemize}

We evaluate the performance in a similar way as in  \cite{bb:cvpr_query}: for the Holiday dataset we use the mean Average Precision (mAP) value and for the Ukbench dataset we use the N-S score \textcolor{blue}{because it contains only $4$ correct images for each query}. Table \ref{tab:result_1} shows the results of our experiments. Our method outperforms all baseline algorithms and \cite{bb:state_graph}. The result of the proposed method is further comparable with \cite{bb:cvpr_query}. However, we use only image features from the datasets to run our algorithm, unlike the state-of-the-art method \cite{bb:cvpr_query}, which uses a large Flickr dataset to train the feature distributions. Also, it is worth noting that both datasets under consideration contain only a few ground truth examples for every query. It gives further advantages to the algorithm in \cite{bb:cvpr_query}, which calculates feature weights to get a final result based on information about a query. 

For the Holiday dataset, our method outperforms Baseline 1, Baseline 3 and \cite{bb:state_graph}. It shows that we can achieve improvement using labeled data. Also, our method gives better result than Baseline 2, which has access to labeled data. It shows that the combination of different layers using the neighborhood of every node is more effective than using information about the query node alone. The algorithm in \cite{bb:cvpr_query}, \textcolor{blue}{which is an unsupervised method}, works slightly better, however they use information about feature distribution from a large Flickr collection. \cite{bb:cvpr_query} combines features with weight where one weight is used for whole layer. This strategy is similar to the Baseline 2. 

For the Ukbench dataset we use $k=3$ to construct our kNN graph, because we know that every image in dataset has only four corresponding images. Our method outperforms the baseline and  the state-of-the-art methods. Baseline 1 and Baseline 2 have the same N-S score, which is better than Baseline 3. It happens because, for this dataset, the actual distribution of the features is important. The graph methods give an opportunity to capture this distribution. 

In summary Table \ref{tab:result_1} shows that the results based on the graph methods are very close to the state-of-the-art and our method produces the best results. The results further show the influence of the neighbors nodes to the query image in choosing the right transition probabilities or equivalently in modeling well feature distribution. 

\begin{table}
\centering
\begin{tabular}{|l|c|c|}
  \hline
  Retrieval algorithm&Holidays,mAP\% &Ukbench,N-S score\\
  \hline
  Baseline 1 & 0.7151 & 3.5835\\
  \hline
  Baseline 2 & 0.7214 & 3.5835 \\
  \hline
  Baseline 3 & 0.7193  & 3.5485 \\ 
   \hline
  \cite{bb:cvpr_query} & \bf{0.7580} & 3.5864 \\
  \hline
  \cite{bb:state_graph} &  0.6593 & 3.2625\\
  \hline
  Proposed method& 0.7435 & \bf{3.5899} \\
   \hline

\end{tabular}
\caption{Comparison of our and state-of-the-art algorithms on the Holiday and the Ukbench datasets.}
\label{tab:result_1}
\end{table}

\section{Conclusion}
\label{sec:con}
This work is dedicated to the timely but challenging problem of image retrieval. It tries to mitigate the issues induced by the so-called semantic gap by properly combining multimodal features for image ranking. Currently, researchers use very complicated techniques to solve this problem in image retrieval. We rather show in this paper that combining features of different modalities in a proper way with a multi-layer graph permits to achieve effective retrieval with a simple random walk algorithm. In particular, the proposed solution achieves good image ranking results compared to the state-of-the-art methods. 

We firmly believe that flexible structures like graphs offer promising solutions to capture the underlying geometry of multi-view data. This is confirmed by the performance of our algorithm. We therefore plan to investigate in the future new graph-based methods that properly exploit the data structure in large-scale retrieval problems.

\bibliographystyle{abbrv}
\bibliography{sigproc}  

\end{document}